\documentclass[twocolumn]{aastex631}

\usepackage{inputenc}

\usepackage{color, amssymb, amsmath, natbib, color} 
\usepackage{multirow, placeins, subfigure}
\usepackage{graphicx}
\usepackage{epstopdf}
\usepackage{bm}
\usepackage{overpic}
\usepackage{hyperref}

\definecolor{lgray}{gray}{0.3}
\definecolor{orange}{rgb}{1,0.5,0}

\newdimen\digitwidth
\setbox1=\hbox{0}
\digitwidth=\wd1
\catcode`"=\active
\def"{\kern\digitwidth}

\begin{document}


\title{Split invariant curves in rotating bar potentials}


\author{Tian-Ye Xia}
\affiliation{Department of Astronomy, School of Physics and Astronomy, Shanghai Jiao Tong University\\ 
800 Dongchuan Road, Shanghai 200240, China}
\affiliation{Key Laboratory for Particle Astrophysics and Cosmology (MOE) / Shanghai Key Laboratory for Particle Physics and Cosmology\\
Shanghai 200240, China}
\affiliation{Lanzhou University\\
298 Tianshui Road, Lanzhou, Gansu Province 730000, China}

\author{Juntai Shen}
\affiliation{Department of Astronomy, School of Physics and Astronomy, Shanghai Jiao Tong University\\ 
800 Dongchuan Road, Shanghai 200240, China}
\affiliation{Key Laboratory for Particle Astrophysics and Cosmology (MOE) / Shanghai Key Laboratory for Particle Physics and Cosmology\\
Shanghai 200240, China}
\affiliation{Shanghai Astronomical Observatory, Chinese Academy of Sciences, 80 Nandan Road, Shanghai 200030, China}

\email{jtshen@sjtu.edu.cn}

\begin{abstract}

Invariant curves are generally closed curves in the Poincar\'{e}'s surface of section.  Here we study an interesting dynamical phenomenon, first discovered by \citet{Binney85} in a rotating Kepler potential, where an invariant curve of the surface of section can split into two disconnected line segments under certain conditions, which is distinctively different from the islands of resonant orbits. We first demonstrate the existence of split invariant curves in the Freeman bar model, where all orbits can be described analytically. We find that the split phenomenon occurs when orbits are nearly tangent to the minor/major axis of the bar potential. Moreover, the split phenomenon seems ``necessary'' to avoid invariant curves intersecting with each other.  Such a phenomenon appears only in rotating potentials, and we demonstrate its universal existence in other general rotating bar potentials. It also implies that actions are no longer proportional to the area bounded by an invariant curve if the split occurs, but they can still be computed by other means.

\end{abstract}

\keywords{galaxies, kinematics and dynamics, galaxies structure}

\section{Introduction}

The surface of section (SoS) was first introduced by \citet{book2}, which is the cross section of the 4D phase space cut at $x=0$ with $\dot{x}>0$. The morphology of the surface of section is affected by integrals of motion, which are any function of the phase space coordinates and constant along an orbit \citep{book1}. Regular orbits of the same energy form invariant curves in the surface of section because of the existence of the second/third integral of motion in addition to the energy and the angular momentum. Irregular (chaotic) orbits, which have no additional integrals of motion, distribute as discrete points within the region bounded by the zero velocity curve \citep{book1}.

Rotating and non-rotating potentials also show different properties in the surface of section. In the non-rotating case, the surface of section always has the left-right symmetry. However, it is not true for the surface of section of a rotating potential, where the Jacobi energy, a combination of energy and angular momentum in the inertial frame ($E_{\rm{J}}=E-\bm{\Omega}\cdot\bm{L}$), becomes the only classical integral of motion.

Invariant curves are generally closed curves surrounding a periodic parent orbit. Since an orbit remains on such a closed curve, we then know that there must be an additional integral of motion besides classical integrals of motion.

However, there exist some invariant curves which split into two disconnected line segments under certain conditions.  \citet{Binney85} first noticed this phenomenon for a rotating Kepler potential when they demonstrated that the phase-space volume is not necessarily proportional to the area within an invariant curve unless the recurrence time is fixed. While the action cannot be calculated simply by the area of an invariant curve when the split occurs, one can still turn to line integrals to complete the calculations \citep{Binney85}.

The split invariant curves differ distinctively from the  disconnected islands of resonant orbits in both the appearance and location of the invariant curves in the SoS. Thus, we find this split phenomenon interesting and worth exploring further. 

We first study the properties of split invariant curves of the planar Freeman bar potential. The Freeman bar model was proposed by \citet{freeman1}, and all the orbits in this model have analytical expressions. It is simple yet illustrative, thus it is often used to draw general conclusions of rotating bar potential. The model differs from other more realistic bar models mainly on two aspects: Firstly, it does not have a co-rotation radius; Secondly, it contains only regular orbits, and has no chaotic orbits.

In $\S$\ref{section:Freeman} we briefly review the results of the Freeman bar model. In $\S$\ref{section:SoS} we introduce split invariant curves in the Freeman bar model and study their properties. In $\S$\ref{sec:discussion} we demonstrate the existence of split invariant curves in other general bar potentials, such as a logarithmic potential. We conclude in $\S$\ref{section:conclusion}.


\section{Analytical Expression of orbits in the Freeman bar model}
\label{section:Freeman}
The Freeman bar is a razor-thin elliptical disk rotating with an angular speed of $\Omega_{\rm{b}}$ \citep{freeman1}. The boundary of the disk is elliptical (${x^2}/{a^2}+{y^2}/{b^2}=1$). The surface density of the disk is \citep{freeman2}:
\begin{equation}
 \Sigma(x,y)=\Sigma_0\sqrt{1-\frac{x^2}{a^2}-\frac{y^2}{b^2}}.
\end{equation}

The quadratic potential is given by \cite{freeman2}:

\begin{equation}
 \Phi(x,y)=\frac{1}{2}(\Omega_x^{2}x^{2}+\Omega_y^{2}y^{2}),
\end{equation}

where $\Omega_x/\Omega_y$ is a function of $b/a$ through complete elliptical integrals \citep{freeman2}.

The equation of motion is:
\begin{equation}
    \bm{\ddot{r}}=-\bm{\nabla}\Phi_{\rm{eff}}-2\bm{\Omega_{\rm{b}}}\times\bm{\dot{r}}.
\end{equation}
where $\Phi_{\rm{eff}}$ is the effective potential of the Freeman bar:

\begin{equation}
\begin{aligned}
\Phi_{\rm{eff}}&=\frac{1}{2}\Omega_x^{2}x^{2}+\frac{1}{2}\Omega_y^{2}y^{2}-\frac{1}{2}\Omega_{\rm{b}}^{2}(x^{2}+y^{2})\\
&=\frac{1}{2}\Phi_{xx}{x^{2}}+\frac{1}{2}\Omega_{yy}{y^{2}},
\end{aligned}
\end{equation}
where $\Phi_{xx}=\Omega_x^{2}-\Omega_{\rm{b}}^{2}$ and $\Phi_{yy}=\Omega_y^{2}-\Omega_{\rm{b}}^{2}$.

The equation of motion can also be written as:

\begin{equation}
\begin{aligned}
&\ddot{x}=2\Omega_b\dot{y}-\Phi_{xx}x,\\
&\ddot{y}=2\Omega_b\dot{x}-\Phi_{yy}y.
\end{aligned}
\end{equation}

The solutions of the above differential equations are \citep{freeman1}:

\begin{equation}
\begin{aligned}
&x=X_{\alpha}\cos(\alpha{t}+\phi_{\alpha})+X_{\beta}\cos(\beta{t}+\phi_\beta),\\
&y=Y_{\alpha}\sin(\alpha{t}+\phi_{\alpha})+Y_{\beta}\sin(\beta{t}+\phi_\beta),
\end{aligned}
\label{equation:solution}
\end{equation}
where $\alpha$ and $\beta$ are two positive roots of equation $x^4-x^2(\Phi_{xx}+\Phi_{yy}+4\Omega_{\rm{b}}^2)+\Phi_{xx}\Phi_{yy}$ and $\alpha<\beta$.

Each orbit in a Freeman bar model is the superposition of two elliptic motions, which are named the $\alpha$-motion and the $\beta$-motion, respectively. The $\alpha$-motion rotates with a lower frequency $\alpha$ in the same direction of the co-rotating frame, whereas the $\beta$-motion rotates with a higher frequency $\beta$ in the retrograde direction. The axial ratios $q_\alpha$ and $q_\beta$ of the $\alpha$-motion and the $\beta$-motion are constants:
\begin{equation}
\begin{split}
q_\alpha=\frac{Y_{\alpha}}{X_{\alpha}}=\frac{\Phi_{xx}-\alpha^2}{2\Omega_{\rm{b}}\alpha}=\frac{2\Omega_{\rm{b}}\alpha}{\Phi_{yy}-\alpha^2},\\
q_\beta=\frac{Y_{\beta}}{X_{\beta}}=\frac{\Phi_{xx}-\beta^2}{2\Omega_{\rm{b}}\beta}=\frac{2\Omega_{\rm{b}}\beta}{\Phi_{yy}-\beta^2},
\end{split}
\end{equation}

The $\alpha$-motion is prograde and the $\beta$-motion is retrograde. The orbit becomes the periodic $x_4$ orbit when $X_{\alpha}=0$, and becomes the periodic $x_1$ orbit when $X_{\beta}=0$ in the terminology of bar orbital families by \citet{contopoulos}.

The Jacobi energy is conserved during the motion. It can be written as:

\begin{equation}
\begin{aligned}
E_{\rm{J}}&=\frac{1}{2}({\vert{\dot{x}}\vert}^2+{\vert{\dot{y}}\vert}^2)+\Phi_{\rm{eff}}\\
&=\frac{1}{2}X_\alpha^2(\alpha^2+\Phi_{yy}q_\alpha^2)+\frac{1}{2}X_\beta^2(\beta^2+\Phi_{yy}q_\beta^2).
\end{aligned}
\end{equation}

There is a second integral of motion in addition to $E_{\rm{J}}$. \cite{freeman1} showed that the Hamiltonian of the system could be separated through the canonical coordinate transformation. Thus, the analytical expression of $I_2$ may be obtained as \citep{freeman1}:

\begin{equation}
I_2=\frac{1}{2}q_{\alpha}X_{\alpha}^2\frac{\alpha}{f},
\end{equation}
where $f={\alpha}q_{\alpha}-{\beta}q_{\beta}$.

From the equation above, $I_2$ is proportional to the area of the $\alpha$-motion ($X_{\alpha}^2$). Thus, $X_\alpha$ or $X_\beta$ is the second integral of motion ($I_2$) in addition to a given $E_{\rm J}$.



\section{Split invariant curves in the Freeman bar model}
\label{section:SoS}

\begin{figure}[!htbp]
  \centering
  \begin{overpic}[width=0.42\textwidth]{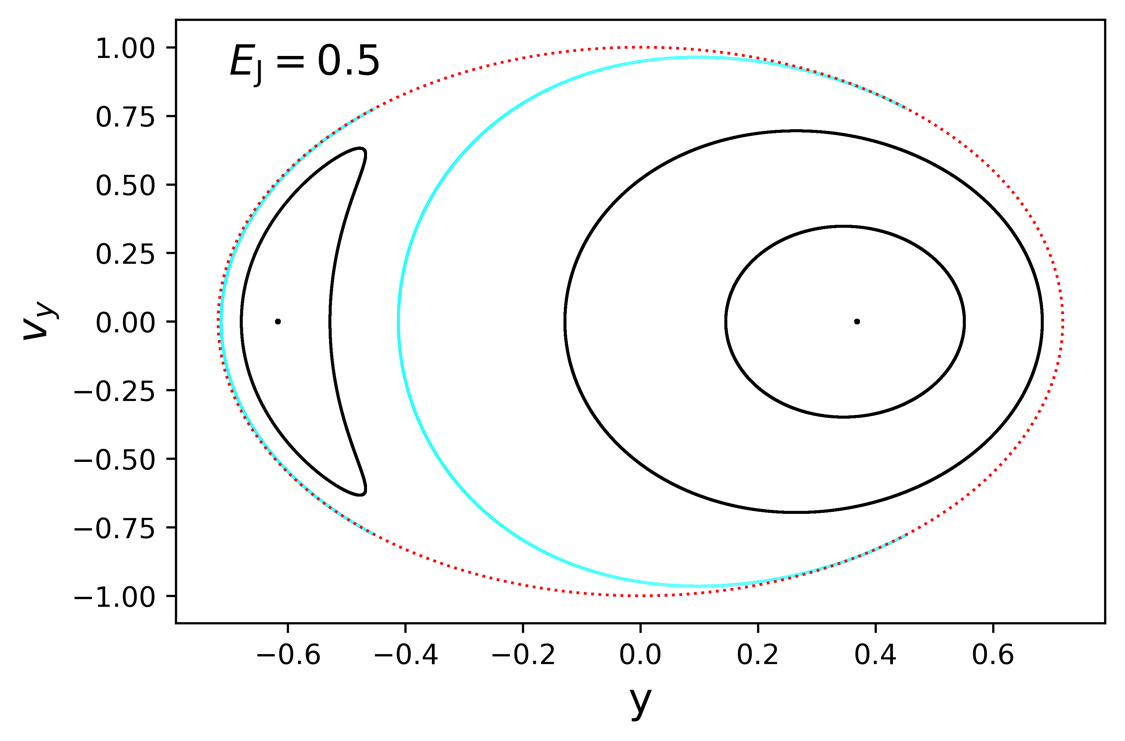}
     \put(1.5,62){\small\bfseries\color{black}{$(a)$}}
  \end{overpic}
  \begin{overpic}[width=0.45\textwidth]{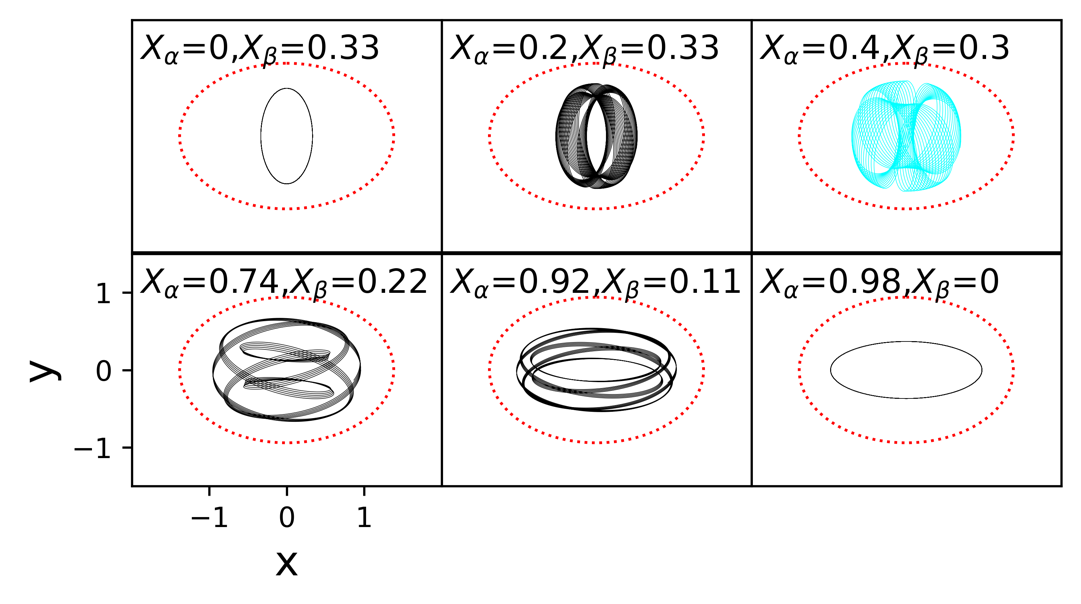}
     \put(2,52){\small\bfseries\color{black}{$(b)$}}
  \end{overpic}
  \begin{interactive}{animation}{3Dps.mp4}
    \begin{overpic}[width=0.45\textwidth]{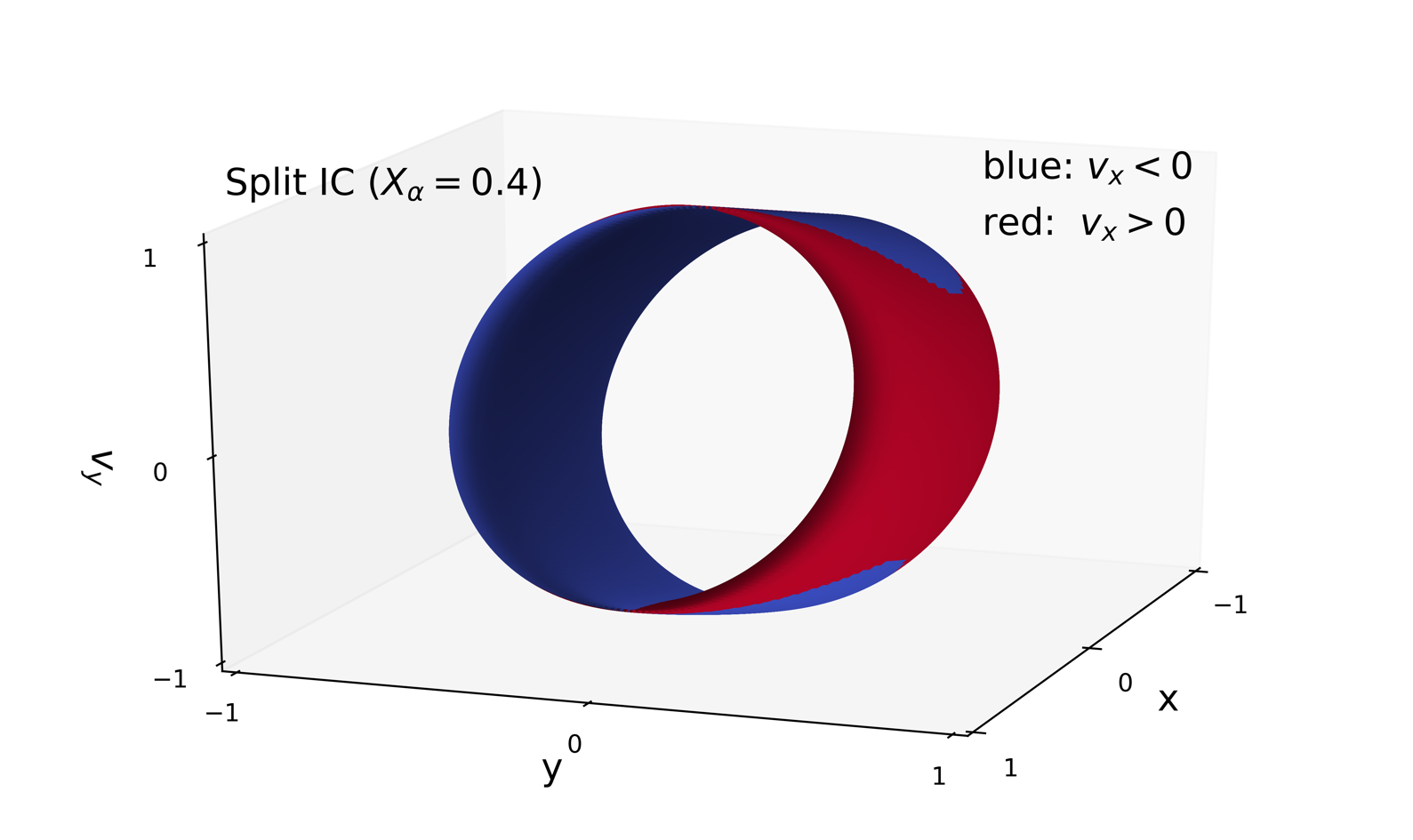}
       \put(2,52){\small\bfseries\color{black}{$(c)$}}
    \end{overpic}
  \end{interactive}
  \caption{
  Phase space properties of a Freeman bar with $\Omega_x^2=1$, $\Omega_y^2=2$ and $\Omega_{\rm{b}}=0.25$ when $E_{\rm{J}}= 0.5$. The boundary of the bar is characterized by $x^2/a^2+y^2/b^2=1$, where $a=1.38$ and $b=0.94$. (a) the surfaces of section of the Freeman bar with the zero velocity curve shown as red dotted curves. Six orbits are selected, covering the lowest $X_\alpha$($X_\alpha=0$) to the highest $X_\alpha$($X_\beta=0$). (b) selected representative orbits corresponding to the invariant curves in (a), where the red dotted curves represent the boundary of the bar. The cyan invariant curve and orbit represent the split invariant curve and its corresponding orbit. (c) the 3D phase space of the cyan split invariant curve ($v_y$ surface as a function of $x$ and $y$). The blue part represents the surface with $v_x < 0$, whereas the red part represents the surface with $v_x > 0$. An animated version of this figure with a $450^{\circ}$ rotation of the viewpoint is also available through the link. The video duration is 11 seconds.}
  \label{fig:SoS}

\end{figure}

Surfaces of section for a Freeman bar with $\Omega_x^2=1$ and $\Omega_y^2=2$ are shown in Figure~\ref{fig:SoS}a.
Note that a SoS records the values of ($y$,  $\dot{y}$) when an orbit crosses the bar minor ($y-$) axis with $\dot{x} < 0$ (to eliminate the sign ambiguity). Each curve in a SoS represents one orbit (Figure~\ref{fig:SoS}b), and all curves share the same $E_{\rm{J}}$.


Figure~\ref{fig:SoS}a shows that there are two islands in the SoS. The left island corresponds to the $x_4$ family, where the central point represents the closed periodic $x_4$ orbit (``parent'' $x_4$), and the right island corresponds to the $x_1$ family, where the central point represents the closed periodic $x_1$ orbit (``parent'' $x_1$).

Since invariant curves can fill the space bounded by the zero velocity curve in the SoS, all orbits in the Freeman bar model are regular.

In general, invariant curves are closed curves in the SoS. However, we find that some invariant curves in the Freeman bar model may split into two disconnected line segments (the cyan curve in Figure~\ref{fig:SoS}a). These orbits are still the same regular orbits because both of the split parts have the same $I_2$.

\begin{figure*}[ht!]
  \centering
  \includegraphics[width=0.85\textwidth]{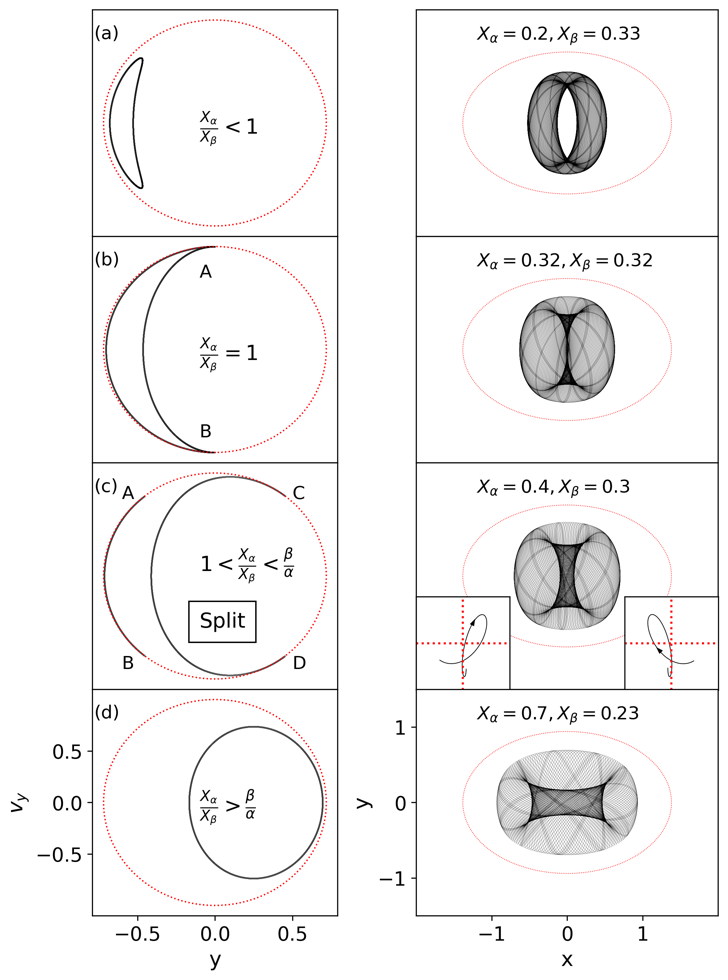}
  \caption{Split invariant curves and their corresponding orbits for a Freeman bar with $\Omega_x^2=1$, $\Omega_y^2=2$ and $\Omega_{\rm{b}}=0.25$ when $E_{\rm{J}}=0.5$. Each row gives the properties of an orbit with given $X_\alpha$ and $X_\beta$. The left column shows surfaces of section with the zero velocity curves shown as the red dotted curves; the right column shows the corresponding orbits. (a) an $x_4$ orbit with its closed, unsplit invariant curve.  (b) an orbit with its unsplit invariant curve, but with two tangent points A and B to the $y$-axis. (c) an orbit with its split invariant curve (the same cyan orbit in Figure~\ref{fig:SoS}), where points A, B, C and D represents four tangent points to the $y$-axis. The left inset in the left panel of (c) shows detailed part of the orbit as it just moves from the left part of the split invariant curve to the right part, whereas the right inset shows the opposite. (d) an $x_1$ orbit with its closed, unsplit invariant curve.}
  \label{fig:split}
\end{figure*}

\subsection{Condition for the split phenomenon in the Freeman bar}
\label{section:condition}

Figure~\ref{fig:split} describes the properties of split and unsplit invariant curves with a given $X_\alpha$ and $X_\beta$, which may replace $I_2$ and $E_{\rm J}$. Figure~\ref{fig:split}a and Figure~\ref{fig:split}b show two unsplit invariant curves. The invariant curve in Figure~\ref{fig:split}b is the critical case where the split phenomenon is just about to happen.

Figure~\ref{fig:split}c shows the detailed properties of the orbit when the split phenomenon occurs. The left parts of these invariant curves are very close to the zero velocity curve, so these orbits cross the $y$-axis with nearly zero $v_x$. Clearly, the $y$-intercepts of the orbits change sign when they jump from one part of the split invariant curve to the other. We therefore deduce that the split invariant curves should correspond to orbits that are nearly tangent to the $y$-axis.

The insets in Figure~\ref{fig:split}c illustrate how the orbits change from one side of the split invariant curve to the other in the SoS. \cite{Binney85} attributed such turn-around orbit to the sign change of the value of the varying angular velocity relative to the pattern speed, as shown in Figure~3 in their paper. 

When an orbit is tangent to the $y$-axis, we have:
\begin{equation}
\begin{aligned}
&x=X_{\alpha}\cos{\theta_{\alpha}}+X_{\beta}\cos{\theta_{\beta}}=0,\\
&\dot{x}=-{\alpha}X_{\alpha}\sin{\theta_{\alpha}}-{\beta}X_{\beta}\sin{\theta_{\beta}}=0,
\end{aligned}
\label{equation:velocity}
\end{equation}
where $\theta_\alpha=\alpha{t}+\phi_\alpha$ and $\theta_\beta=\beta{t}+\phi_\beta$.

This condition is equivalent to $L_z=x\dot{y}-y\dot{x}=0$, with $x=0$ and $y\neq{0}$, yet different from the trivial $L_z=0$ as a box orbit passes directly through the origin ($x=y=0$). Note that $L_z$ here is defined as the angular momentum in the co-rotating frame, whereas $L_z$ in \cite{Binney85} is defined in the inertial frame.

The solutions are:

\begin{equation}
\begin{aligned}
&{\cos}^2{\theta_{\alpha}}=\frac{{\alpha}^2m^2-{\beta}^2}{({\alpha}^2-{\beta}^2)m^2},\\
&{\cos}^2{\theta_{\beta}}=\frac{{\alpha}^2m^2-{\beta}^2}{{\alpha}^2-{\beta}^2},\\
&{\sin}^2{\theta_{\alpha}}=\frac{(m^2-1){\beta}^2}{({\beta}^2-{\alpha}^2)m^2},\\
&{\sin}^2{\theta_{\beta}}=\frac{(m^2-1){\alpha}^2}{{\beta}^2-{\alpha}^2},
\end{aligned}
\end{equation}
where $m\equiv{X_{\alpha}}/{X_{\beta}}$ and $m>0$.

According to the range of the trigonometric function, the condition for orbits that are tangent to the $y$-axis is:

\begin{equation}
1\leq{m}\leq\frac{\beta}{\alpha}.
\end{equation}

If $m=1$ (Figure~\ref{fig:split}b), $\sin{\theta_{\alpha}}=0$ and $\sin{\theta_{\beta}}=0$. The $y$-axis values of the tangent points are 0. In this case, there are only two tangent points (points A and B in Figure~\ref{fig:split}b). They are both located on the crescent island in the left, so the invariant curve does not split.

If $1<m<{\beta}/{\alpha}$ (Figure~\ref{fig:split}c), since $X_\alpha$, $X_\beta$, ${\alpha}X_\alpha$, ${\beta}X_\beta$ are positive, $(\cos{\theta_\alpha},\cos{\theta_\beta},\sin{\theta_\alpha},\sin{\theta_\beta})$ has four groups of solutions, which correspond to four tangent points. According to Equations~\ref{equation:solution} and \ref{equation:velocity}, $y$ and $\dot{y}$ can be written as:
\begin{equation}
\begin{aligned}
y&=Y_{\alpha}\sin{\theta_{\alpha}}+Y_{\beta}\sin{\theta_{\beta}}\\
&=\frac{X_\alpha}{\beta}(\beta{q_\alpha}-\alpha{q_\beta})\sin{\theta_\alpha}\propto{\sin{\theta_\alpha}}\\
&=\frac{X_\beta}{\alpha}(\alpha{q_\beta}-\beta{q_\alpha})\sin{\theta_\beta}\propto{\sin{\theta_\beta}},\\
\dot{y}&={\alpha}Y_{\alpha}\cos{\theta_{\alpha}}+{\beta}Y_{\beta}\cos{\theta_{\beta}}\\
&=X_\alpha(\alpha{q_\alpha}-\beta{q_\beta})\cos{\theta_\alpha}\propto{\cos{\theta_\alpha}}\\
&=X_\beta(\beta{q_\beta}-\alpha{q_\alpha})\cos{\theta_\beta}\propto{\cos{\theta_\beta}},\\
\end{aligned}
\end{equation}
so $y$ and $\dot{y}$ are related to $\theta_\alpha$ and $\theta_\beta$. As for the four tangent points, since $\vert{\sin{\theta_\alpha}}\vert$, $\vert{\cos{\theta_\alpha}}\vert$, $\vert{\sin{\theta_\beta}}\vert$, and $\vert{\sin{\theta_\beta}}\vert$ have the same values,  $\vert{y}\vert$ and $\vert\dot{y}\vert$ are constants. Consequently, the four symmetrical tangent points are located in the four quadrants, respectively, and distributed on the left and right islands in pairs (shown as points A, B, C, and D in Figure~\ref{fig:split}c). In this case, the invariant curve splits into two disconnected segments.

If $m={\beta}/{\alpha}$, $\cos{\theta_{\alpha}}=0$ and $\cos{\theta_{\beta}}=0$. The $v_y$ value of the tangent points is 0. There are two tangent points, which are located in the left and right islands, respectively. In this case, the invariant curve splits into a point in the left and a curve in the right (not shown in Figure~\ref{fig:split}).

Figure~\ref{fig:split}d shows the case of $m>{\beta}/{\alpha}$, where the invariant curve does not split.

In summary, the condition for the occurrence of split invariant curves is:

\begin{equation}
1<\frac{X_{\alpha}}{X_{\beta}}\leq\frac{\beta}{\alpha},
\label{equation:condition}
\end{equation}
where
\begin{equation}
\frac{\beta}{\alpha}=\sqrt{1+\frac{2\sqrt{\gamma^2-4\Phi_{xx}\Phi_{yy}}}{t-\sqrt{\gamma^2-4\Phi_{xx}\Phi_{yy}}}},
\end{equation}
and
\begin{equation}
\gamma=\Phi_{xx}+\Phi_{yy}+4\Omega_{\rm{b}}^2.
\end{equation}

\begin{figure*}[htbp!]
  \centering
  \includegraphics[width=0.44\textwidth]{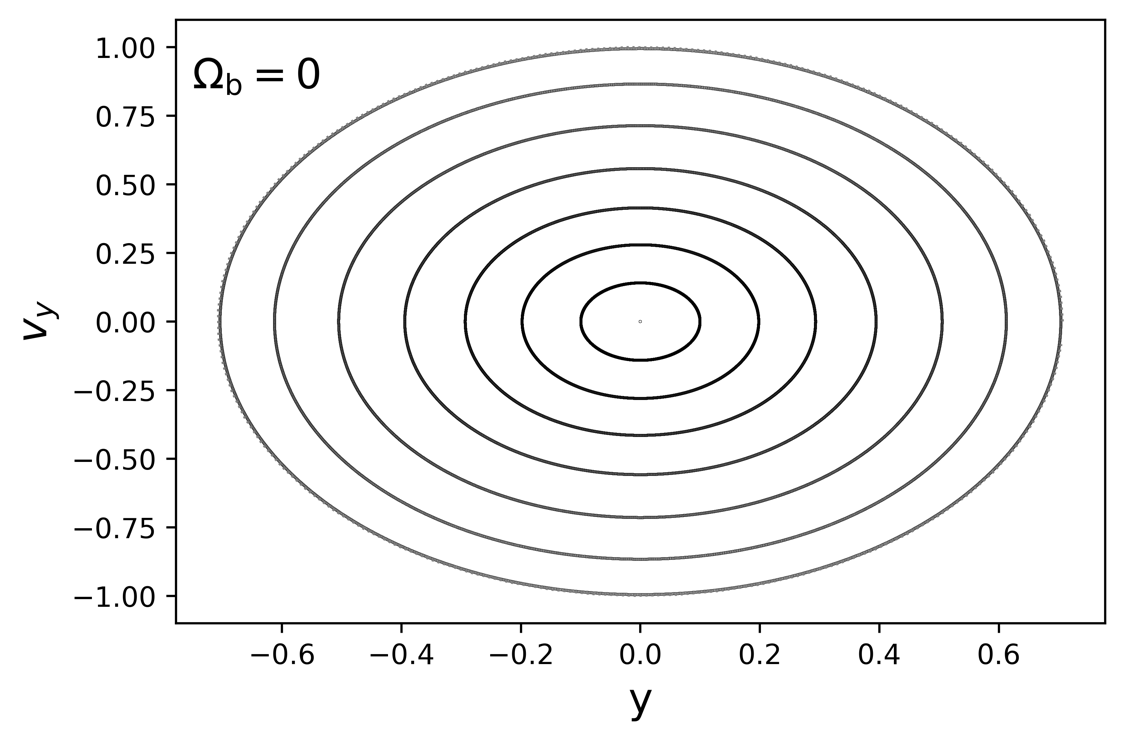}
  \includegraphics[width=0.44\textwidth]{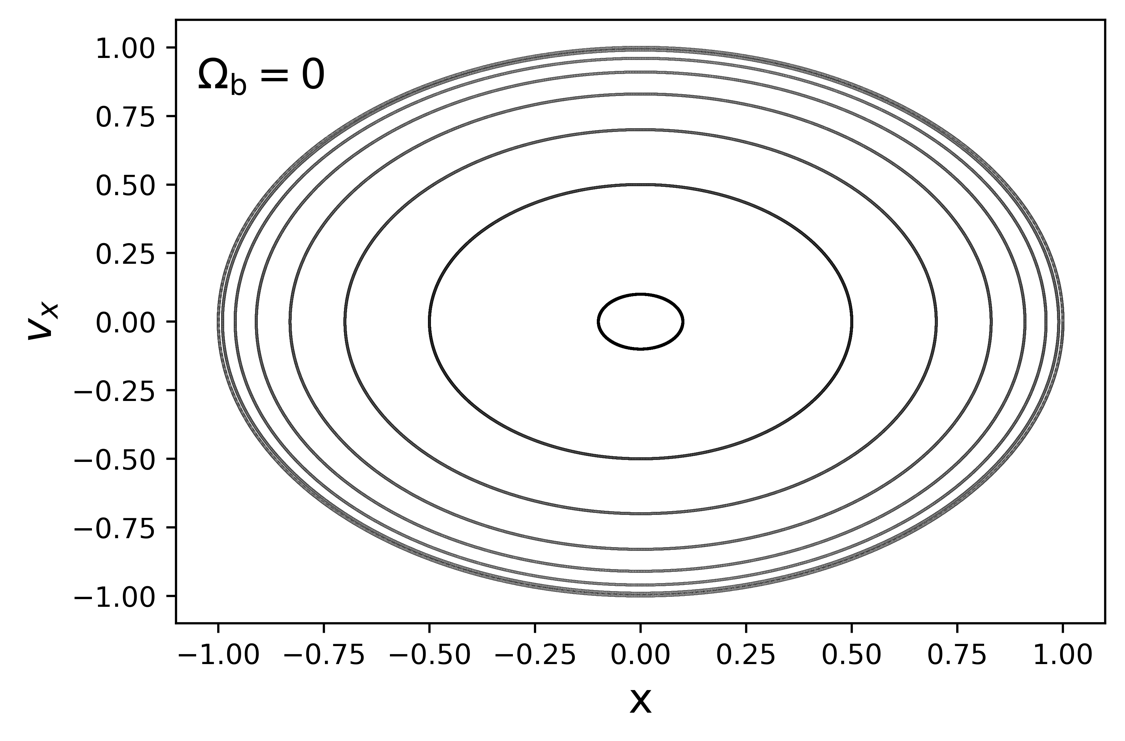}\\
  \includegraphics[width=0.44\textwidth]{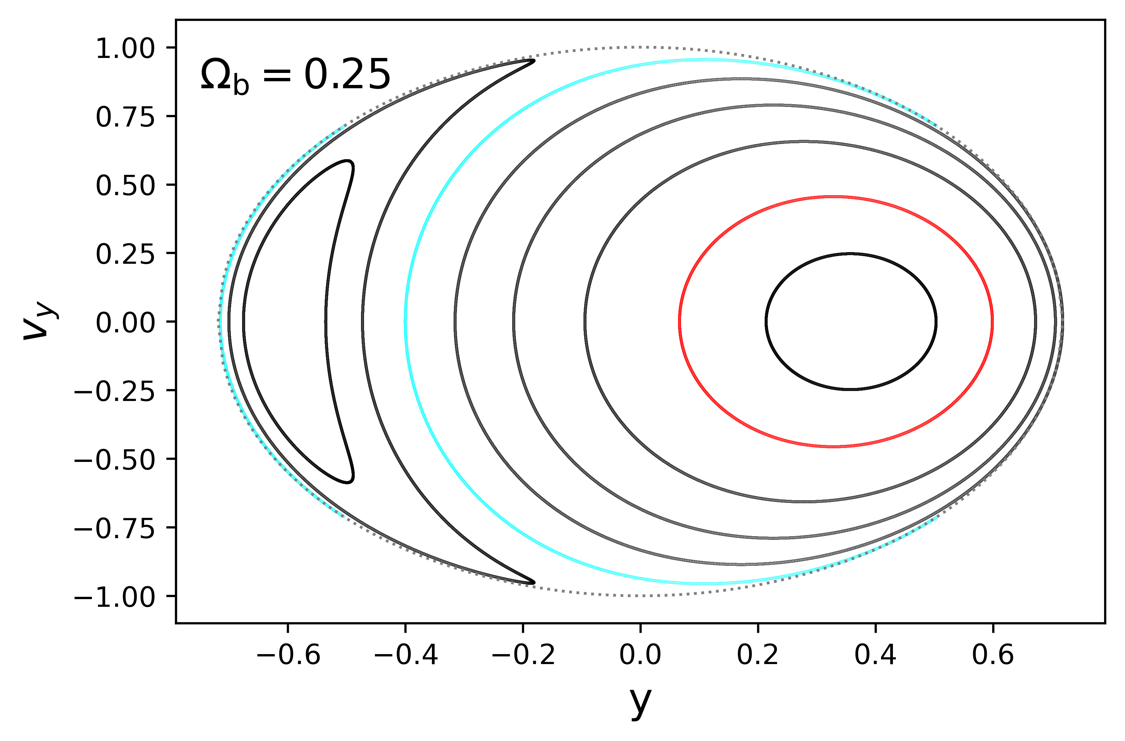}
  \includegraphics[width=0.44\textwidth]{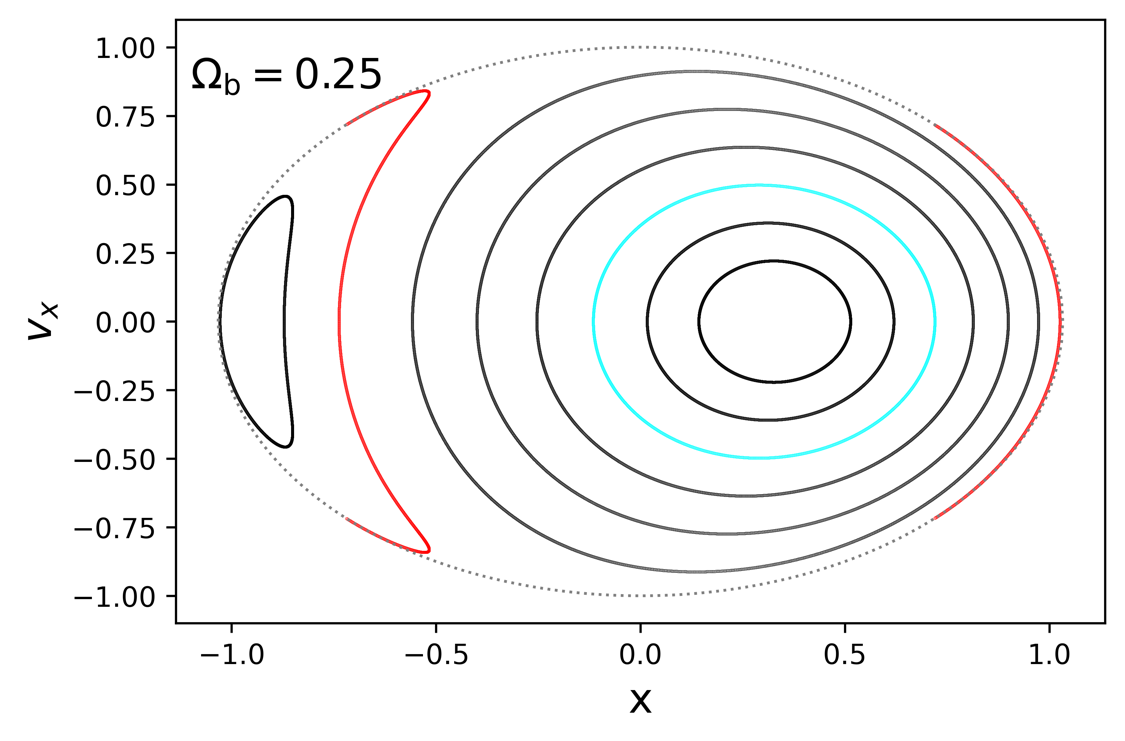}\\
  \includegraphics[width=0.44\textwidth]{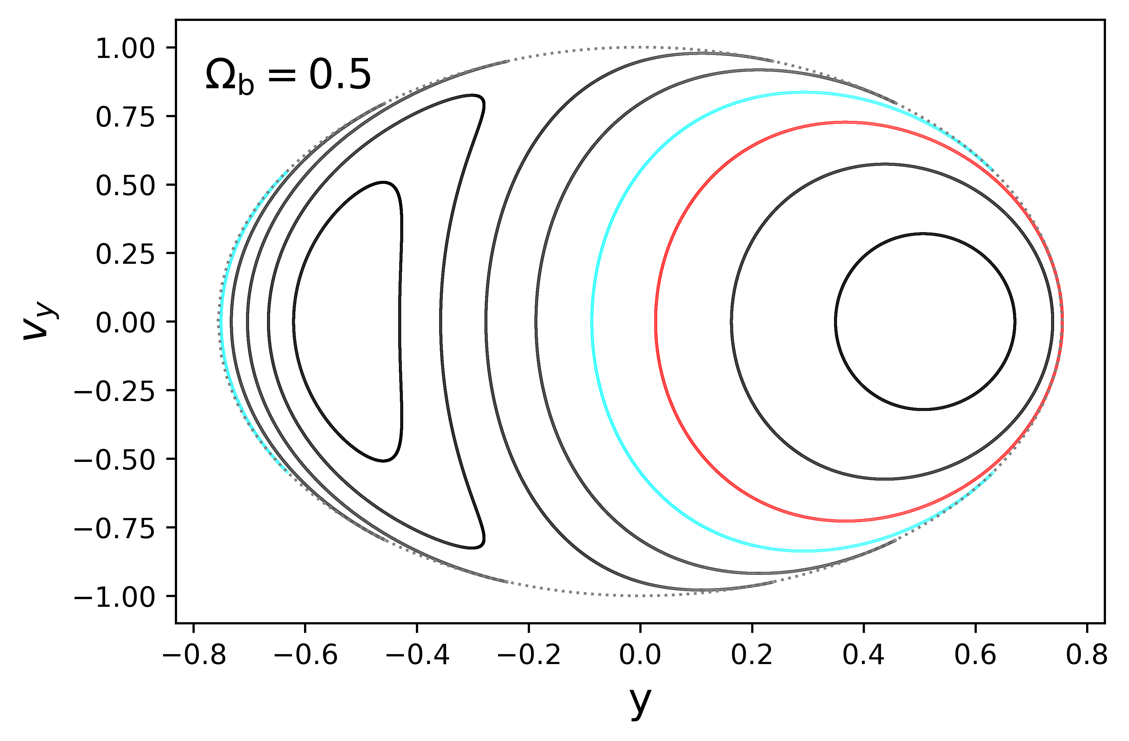}
  \includegraphics[width=0.44\textwidth]{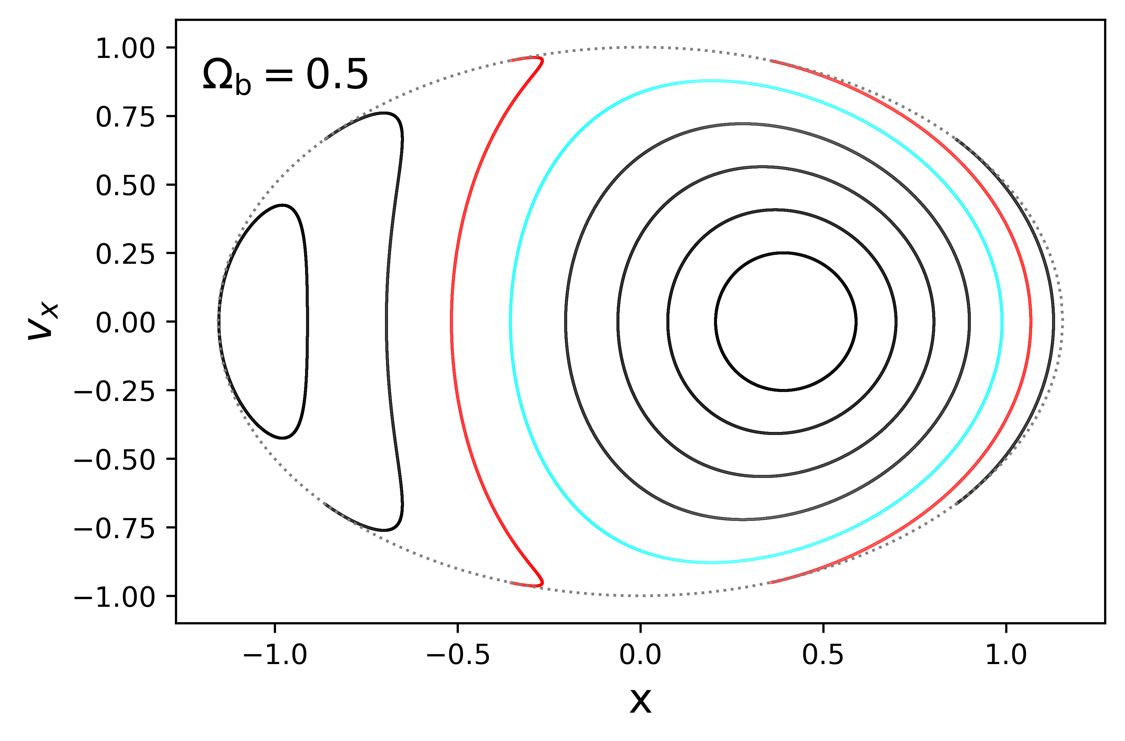}\\
  \includegraphics[width=0.44\textwidth]{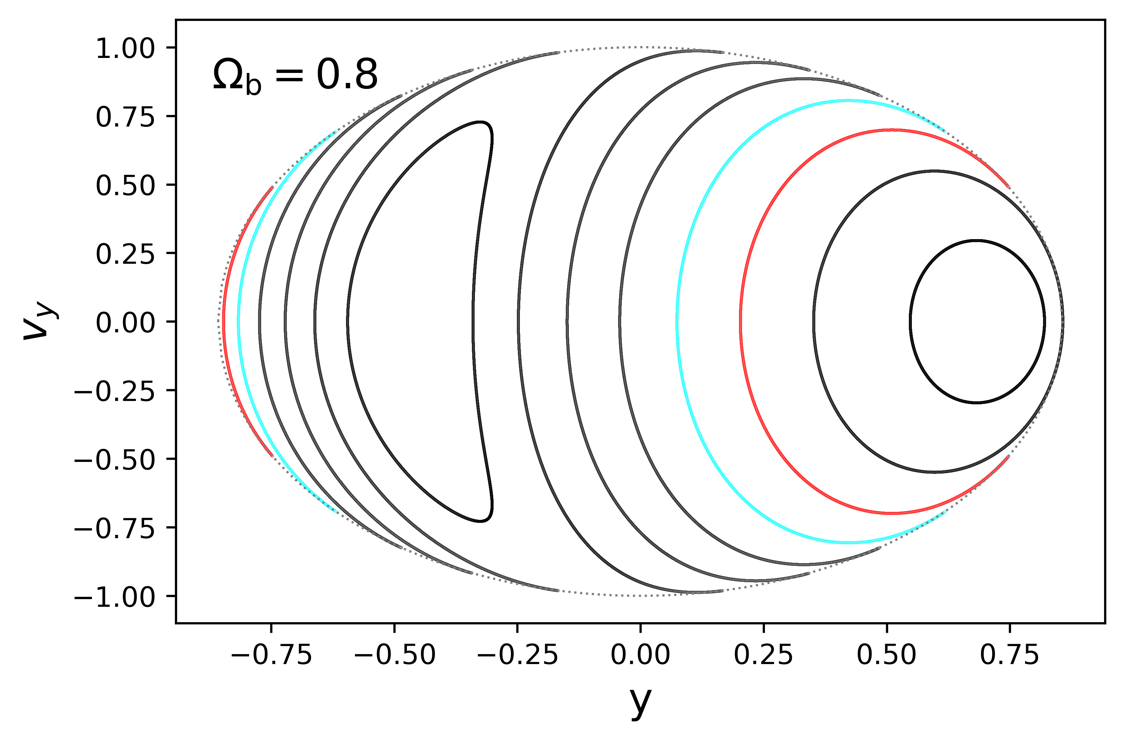}
  \includegraphics[width=0.44\textwidth]{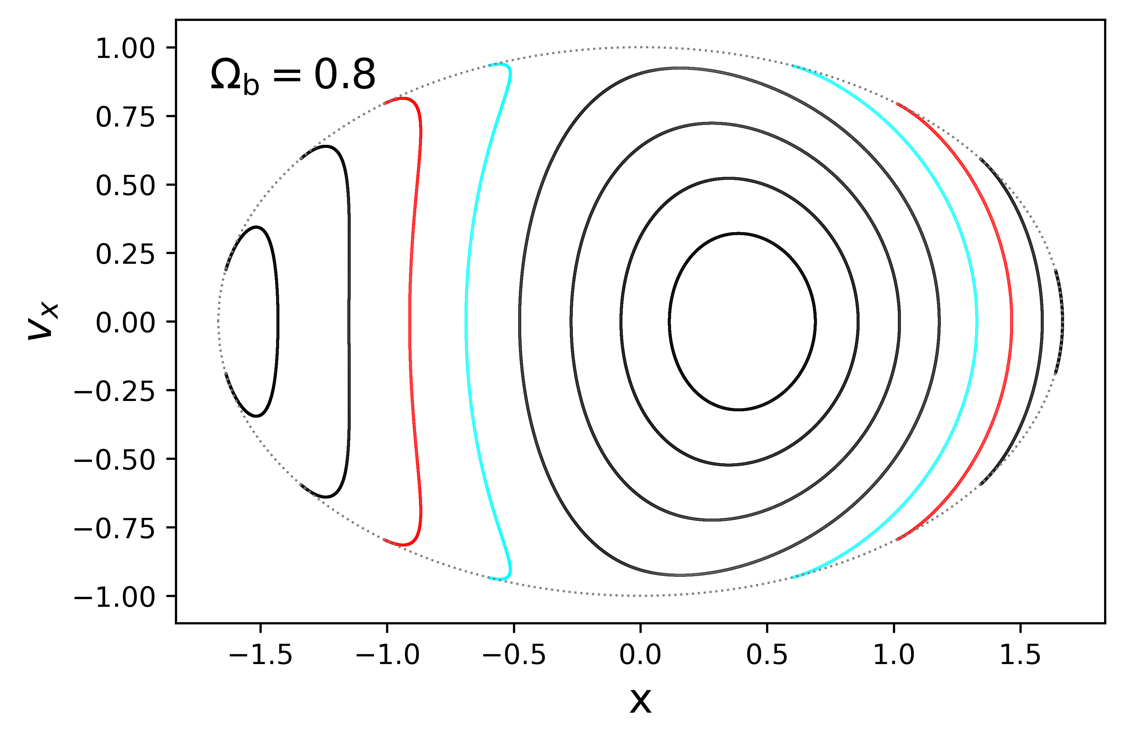}\\
  \caption{Surfaces of section in the $y-v_y$ and $x-v_x$ sections for Freeman bars with $\Omega_x^2=1$, $\Omega_y^2=2$ and different $\Omega_{\rm{b}}$ when $E_{\rm{J}}=0.5$. The grey curves represent the zero velocity curve. For each Freeman bar, the red and cyan curves are the representative split invariant curves in the $y-v_y$ or $x-v_x$ section. Invariant curves which have the same color (except the black curves) for the both $y-v_y$ and $x-v_x$ sections correspond to the same orbit. For the cases with $\Omega_{\rm{b}}=0$, no split invariant curve exists. For the cases with $\Omega_{\rm{b}}=0.25$ and $\Omega_{\rm{b}}=0.5$, the split invariant curves (the cyan curves) which split in the $y-v_y$ section do not split in the $x-v_x$ section, and the red curves show vice versa. For the case with $\Omega_{\rm{b}}=0.8$, the cyan and red invariant curves split both in $y-v_y$ and $x-v_x$ sections.}
  \label{fig:varyomegab}
\end{figure*}

From the above equations, one may also notice that the condition of split invariant curves is related to $\Omega_{\rm{b}}$. The range of the condition increases monotonically with the increasing $\Omega_{\rm{b}}$. Consequently, the split phenomenon is more pronounced in more rapidly rotating systems (see the left column of Figure~\ref{fig:varyomegab}). We also find that no split invariant curve exists in a non-rotating potential, but there are always split invariant curves in rotating ones (also see \citealt{Binney85}).

\subsection{3D phase space of split invariant curves}

We further explore the properties of the 3D phase space of split invariant curves. The phase space of the cyan split invariant curve in Figure~\ref{fig:SoS}a forms a tilted torus in the 3D space of $x$, $y$ and $v_y$ at a given $E_{\rm J}$ (Figure~\ref{fig:SoS}c). We also include an animation illustrating a $450^{\circ}$ rotation of the viewpoint of the torus in Figure~\ref{fig:SoS}c's link. The blue part in the video represents the surface with $v_x < 0$, whereas the red part represents the surface with $v_x > 0$. We find that the split phenomenon is strongly affected by the junction of the surface with $v_x < 0$ and $v_x > 0$. The $v_x < 0$ requirement forces the SoS to split for some orbits.

In Figure~\ref{fig:SoS}c and the animation, we focus on the two small blue patches in the red background with $y>0$ and the two small red patches in the blue background with $y<0$. As $X_\alpha$ increases (orbits transition from $x_1$ to $x_4$ families), the distance between the two small patches of the same color decreases with the $x$-direction width of the whole torus increasing. The split invariant curves begin to appear when the $x=0$ plane cuts through the red patches, and disappear when the two blue patches touch each other. Thus, the size and position of the small patches determine whether or not the split invariant curves appear.



\section{Discussions} \label{sec:discussion}

\begin{figure}[htb]
  \centering
  \includegraphics[width=0.44\textwidth]{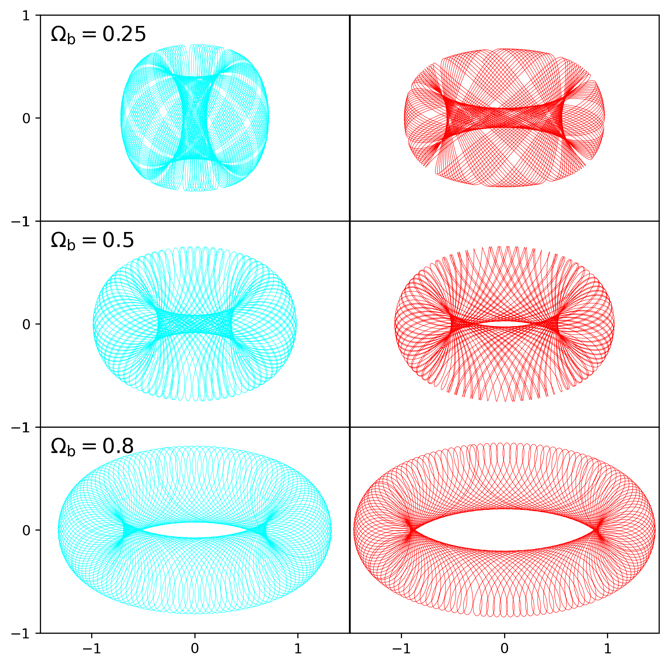}
  \caption{Orbits corresponding to the red and cyan split invariant curves in Figure~\ref{fig:varyomegab} when $\Omega_b=0.25$, $0.5$, and $0.8$.}
  \label{fig:omega_or}
\end{figure}

Figure~\ref{fig:varyomegab} also shows that split invariant curves exist in the $x-v_x$ section of SoS, the cross section of the phase space cut at $y=0$ with $\dot{y}<0$, besides the $y-v_y$ section on which we have focused as of now. When the bar pattern speed is relatively small ($\Omega_b=0.25$ and $\Omega_b=0.5$), the invariant curve that splits in the $y-v_y$ section does \emph{not} split in the $x-v_x$ section, and vice versa. Intriguingly, when the bar pattern speed is large enough ($\Omega_b=0.8$), there exist split invariant curves in both the $x-v_x$ and $y-v_y$ sections, corresponding to the same orbit. The orbits of the red and cyan split invariant curve in Figure~\ref{fig:varyomegab} are shown in Figure~\ref{fig:omega_or}.

Similar to the discussions in $\S$\ref{section:condition}, the condition for split invariant curves in the $x-v_x$ section of SoS is that the orbit becomes nearly tangent to the $x$-axis. A similar derivation yields the condition for splitting in the $x-v_x$ section of SoS:

\begin{equation}
-\frac{q_\beta}{q_\alpha}<\frac{X_{\alpha}}{X_{\beta}}\leq-\frac{{\beta}q_\beta}{{\alpha}q_\alpha}.
\label{equation:condition2}
\end{equation}

Note that $q_\beta<0$, and $\alpha\leq\beta$.

\begin{figure}[htb]
  \centering
  \includegraphics[width=0.44\textwidth]{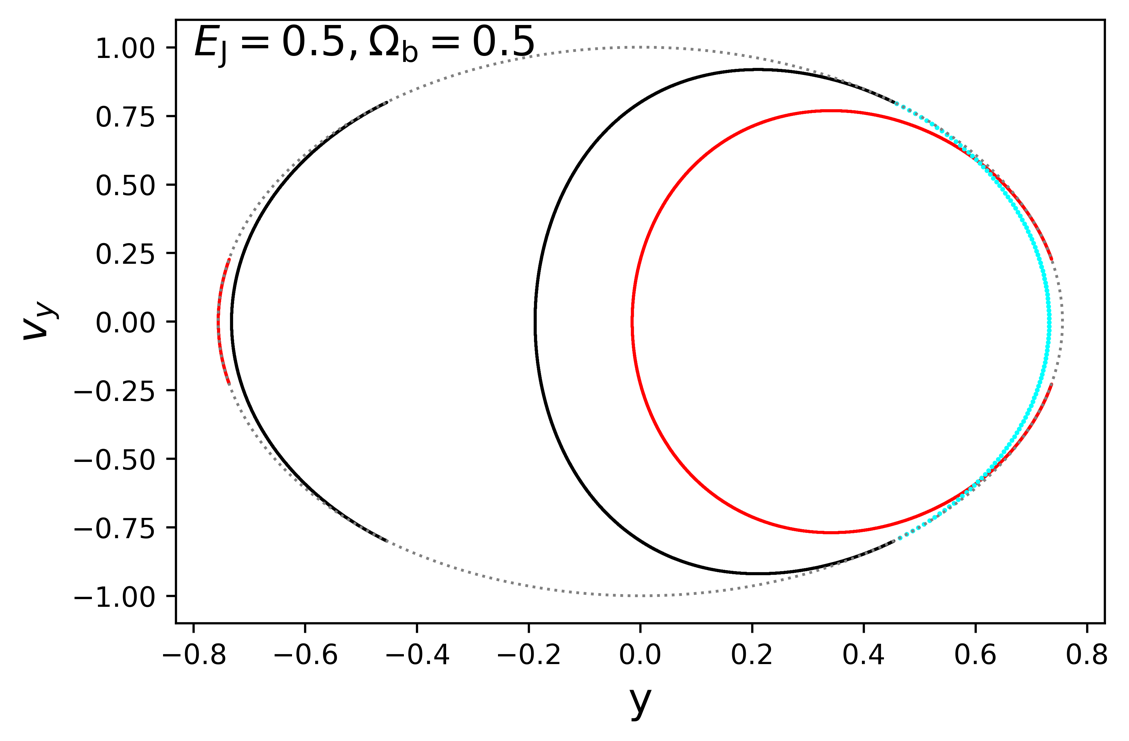}
  \caption{Surfaces of section for a Freeman bar with $\Omega_x^2=1$, $\Omega_y^2=2$ and $\Omega_{\rm{b}}=0.5$ when $E_{\bm{J}}=0.5$. The grey curve represents the zero velocity curve. The black curve and the red curve are the invariant curves when $X_\alpha=0.55$ and $X_\alpha=0.75$, respectively. The cyan curve is the mirror-symmetrized curve of the left part of the black invariant curve if it were not split. }
  \label{fig:cross}
\end{figure}

If the left part of the split invariant curve is mirror-symmetrized to the right side with respect to the $y=0$ line, the mirror-symmetrized curve forms a closed curve with the right part of the invariant curve (the cyan curve in Figure~\ref{fig:cross}). \emph{If the invariant curve were not split, the closed curve would have intersected with another invariant curve} (the red curve in Figure~\ref{fig:cross}), which ought not to happen. In other words, the split phenomenon is ``necessary'' to avoid invariant curves intersecting with each another.

\begin{figure*}[htb]
  \centering
  \includegraphics[width=0.44\textwidth]{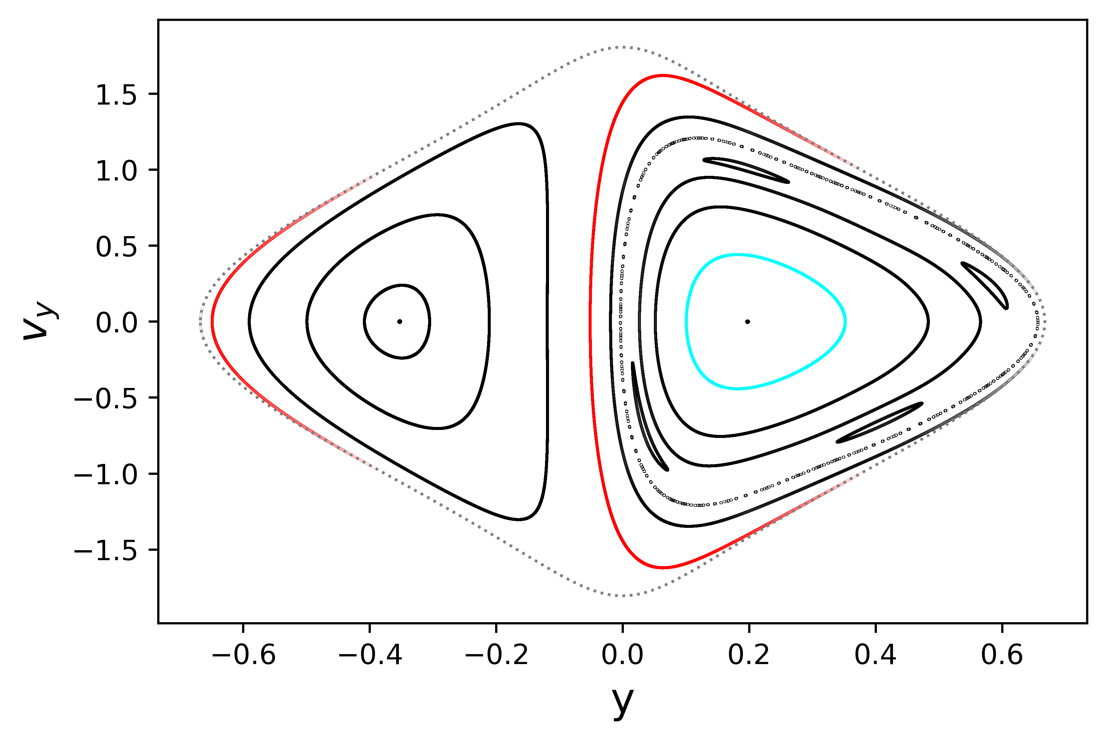}
  \includegraphics[width=0.44\textwidth]{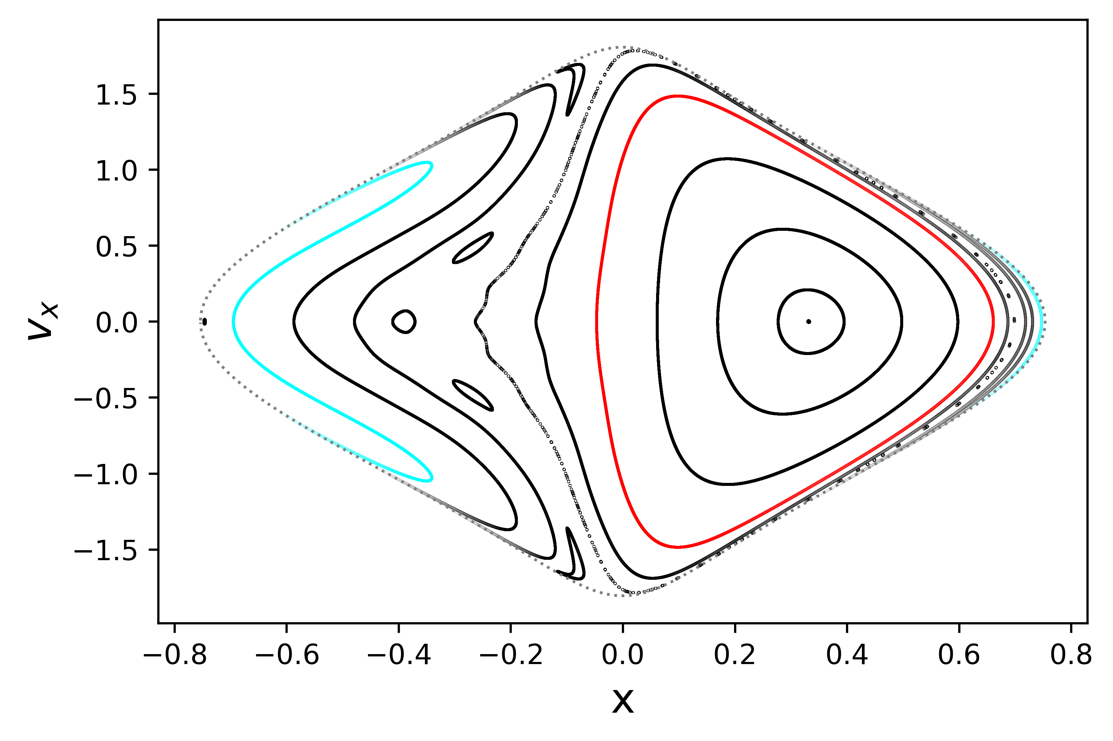}
  \caption{Surfaces of section for a logarithmic potential with $\Omega_{\rm{b}}=0.5$ when $E_{\rm{J}}=-0.337$. The potential is given by $\Phi=\frac{1}{2}\ln{(R_c^2+x^2+y^2/q^2)}$, where $R_c=0.14$ and $q=0.9$. The grey curves represent the zero velocity curve. The cyan and red curves represent the split invariant curves in the $y-v_y$ or $x-v_x$ section. Invariant curves with the same color (except the black curves) in both the $x-v_x$ and $y-v_y$ section correspond to the same orbits.}
  \label{fig:ln}
\end{figure*}

\subsection{The split phenomenon in generic bar potentials}

The Freeman bar potential is quadratic, thus one may wonder if split invariant curves occur only in such an idealized potential. We have tested a more general logarithmic potential that is commonly used:

\begin{equation}
\Phi=\frac{1}{2}\ln{(R_c^2+x^2+y^2/q^2)}, 
\end{equation}
where $R_c=0.14$ and $q=0.9$.

As shown in Figure~\ref{fig:ln}, similar split invariant curves also exist in such a more general bar potential (cyan and red curves).

\subsection{Differences between split invariant curves and disconnected islands formed by resonant orbits}

One may also wonder whether or not the split invariant curves are related to the disconnected islands caused by resonant orbits. However, we think that split invariant curves differ distinctively from those disconnected islands. Firstly, the disconnected islands are always closed ``curves'', not disconnected line segments. Secondly, resonance usually marks the separation between chaotic and regular orbits, while the cyan split invariant curve in Figure~\ref{fig:ln} occurs even between the regular orbits belonging to the same orbital family (also Figure~\ref{fig:varyomegab}), along with the clearly disconnected islands formed by resonant orbits. Moreover, The Freeman bar is an analytical model without chaos or resonant orbits, thus split invariant curves may not be related to resonant orbits.

\section{Conclusions}
\label{section:conclusion}

While invariant curves are generally closed curves in the surface of section, \citet{Binney85} first noticed the existence of split invariant curves in a rotating Kepler potential. In this work, we study the properties of such split invariant curves in a Freeman bar, where all the orbits can be described analytically. We confirm that some invariant curves in a rotating potential split into two disconnected parts under certain conditions, which differs from the disconnected islands formed by resonant orbits. We show that the orbits are nearly tangent to $y$-axis (or $x$-axis), or $L_z =0$ with $y \ne 0$ (or $x \ne 0$), when the phenomenon occurs. The analytical condition for its occurrence is presented in Equations~\ref{equation:condition} and \ref{equation:condition2}. Furthermore, by plotting the 3D phase space of the split invariant curves, we prove that the existence of split invariant curves is the result of the imposed requirement ($v_x < 0$) of the SoS for a rotating potential. We find that the split invariant curves appear in the both $x-v_x$ and $y-v_y$ sections of SoS, but occur only in rotating potentials. We also confirm the existence of the split phenomenon in more generic potentials, such as a logarithmic potential. 

The split phenomenon shows that actions are no longer proportional to the area bounded by an invariant curve if the split occurs, but they can still be computed by other means \citep{Binney85}. Although we have worked out how split invariant curves emerge in the Freeman bar, some questions still remain unsolved, such as the additional implications of split invariant curves, and the properties of the split phenomenon in the presence of addition orbital families other than $x_1$ and $x_4$. Thus, more future studies are needed to better understand this phenomenon and its implications.

\begin{acknowledgements}
We thank the anonymous referee for referring us to Binney et al. (1985) of which we were unaware. We thank Scott Tremaine, Jerry Sellwood, Yu-Jing Qin, Zhi Li for helpful comments. 
The research presented here is partially supported by the National Key R\&D Program of China under grant No. 2018YFA0404501; by the National Natural Science Foundation of China under grant Nos. 12025302, 11773052, 11761131016; by the ``111'' Project of the Ministry of Education of China under grant No. B20019; and by the Chinese Space Station Telescope project. J.S. acknowledges support from a {\it Newton Advanced Fellowship} awarded by the Royal Society and the Newton Fund. This work made use of the Gravity Supercomputer at the Department of Astronomy, Shanghai Jiao Tong University, and the facilities of the Center for High Performance Computing at Shanghai Astronomical Observatory
\end{acknowledgements}

\bibliographystyle{aasjournal}
\bibliography{myrefs}


\end{document}